\documentclass[rnote]{aa} % for the research notes
\def\iphas{\object{IPHAS~J205836.43+503307.2}}
\def\feii{[Fe\,{\sc ii}]}
\def\ki{K\,{\sc i}}
\def\hi{H\,{\sc i}}
\def\ha{H$\alpha$}
\def\hb{H$\beta$}
\def\hg{H$\gamma$}
\def\nii{[N\,{\sc ii}]}
\def\oiii{[O\,{\sc iii}]}
\def\sii{[S\,{\sc ii}]}
\def\siii{[S\,{\sc iii}]}
\def\ariii{[Ar\,{\sc iii}]}
\def\arv{[Ar\,{\sc v}]}
\def\oii{[O\,{\sc ii}]}
\def\oi{[O\,{\sc i}]}
\def\hei{He\,{\sc i}}
\def\heii{He\,{\sc ii}}
\def\fevii{[Fe\,{\sc vii}]}
\def\cav{[Ca\,{\sc v}]}
\def\caii{Ca\,{\sc ii}}
\def\rha{$r'$$-$H$\alpha$}
\def\ri{$r'$$-$$i'$}

\def\jho{($J$$-$$H$)$_0$}
\def\hko{($H$$-$$K_S$)$_0$}
\def\jk{$J$$-$$K_S$}
\def\mincir{\ \raise-2.truept\hbox{\rlap{\hbox{$\sim$}}\raise5.truept \hbox{$<$}\ }}
\def\magcir{\ \raise-3.truept\hbox{\rlap{\hbox{$\sim$}}\raise5.truept \hbox{$>$}\ }}
\def\kms{\relax \ifmmode {\,\rm km\,s}^{-1}\else \,km\,s$^{-1}$\fi}
\def\deg{$^\circ$}
\usepackage{graphicx}
\usepackage{txfonts}
\begin{document}
\title{The new carbon symbiotic star \iphas\thanks{Based 
on observations obtained with the 2.5m~INT and the 4.2m~WHT telescopes 
of the Isaac Newton Group of Telescopes and the 1.5m Carlos Sanchez Telescope, 
operating on the islands of La Palma and Tenerife at the Spanish 
Observatories of the Roque de Los Muchachos and Teide of the Instituto de 
Astrof\'\i sica de Canarias; the 2.1m telescope at San Pedro Martir, Mexico; 
and the GAPC 0.7m Ritchey-Chr\'etien telescope at La Polse di Cougnes, 
Udine, Italy.}}

\author{R.L.M. Corradi\inst{1,2}
           \and
        L. Sabin\inst{3}
         \and
        U. Munari\inst{4}
         \and
        G. Cetrulo\inst{5}
         \and\\
        A. Englaro\inst{5}
         \and
        R. Angeloni\inst{6}
         \and
        R. Greimel\inst{7}
        \and
        A. Mampaso\inst{1,2}
%         \and
%	J.E. Drew\inst{8}
	}
   \offprints{Romano Corradi}
   \institute{
Instituto de Astrof\'\i sica de Canarias, E-38200 La Laguna, 
Tenerife, Spain \email{rcorradi@iac.es}
   \and
Departamento de Astrof\'\i sica, Universidad de La Laguna, 
E-38206 La Laguna, Tenerife, Spain 
\and
Instituto de Astronom\'{i}a, Universidad Nacional Aut\'{o}noma de 
M\'{e}xico, Apdo. Postal 877, 22800 Ensenada, B.C, Mexico
\and 
INAF, Osservatorio Astronomico di Padova, via dell'Osservatorio 8, 
36012 Asiago (VI), Italy
   \and
ANS Collaboration, c/o Osservatorio Astronomico, 36012 Asiago (VI), Italy 
   \and
Departamento de Astronom\'\i a y Astrof\'\i sica, 
Pontificia Universidad Catolica de Chile, Santiago, Chile 
\and
Institut f\"ur Physik, Karl-Franzen Universit\"at Graz, 
Universit\"atsplatz 5, 8010 Graz, Austria
}

\date{Received ... / Accepted ...}

\abstract
{}
{We are performing a search for symbiotic stars using   
IPHAS, the INT \ha\ survey of the northern Galactic plane, 
and follow-up observations.}
{Candidate symbiotic stars are selected on the basis of their IPHAS 
and near-IR colours, and spectroscopy and photometry are obtained to 
determine their nature. We present here observations of the symbiotic star 
candidate \iphas.}
{The optical spectrum shows the combination of a number of emission lines, 
among which are the high-excitation species of \oiii, \heii, \cav, and \fevii, 
and a red continuum with the  features of a star at the cool end 
of the carbon star sequence. 
The nebular component is spatially resolved: the analysis of the
spatial profile of the \nii6583 line in the spectrum indicates a
linear size of $\sim$2$''$.5 along the east-west direction. Its 
velocity structure suggests an aspherical morphology.
The near-infrared excess of the source, which was especially strong in 1999,
indicated that a thick circumstellar dust shell was also present in
the system.
The carbon star has brightened in the last decade by two to four
magnitudes at red and near-infrared wavelengths.  Photometric
monitoring during a period of 60 days from November 2010 to January
2011 reveals a slow luminosity decrease of 0.2~magnitudes.
}
{From the observed spectrophotometric properties and variability, 
we conclude that the source is a new Galactic symbiotic star of the D-type, 
of the rare kind that contains a carbon star, likely a carbon Mira. 
Only two other systems of this type are known in the Galaxy.}
\titlerunning{A new Galactic symbiotic star with a cool carbon star}
\authorrunning{R.L.M. Corradi et al.}
\maketitle

\section{Introduction}

The total population of symbiotic stars in the Galaxy is an important
figure for determining the timescales and characteristics of this binary
channel of stellar evolution. It also has direct implications on the
hypothesis that symbiotic stars are SN Ia progenitors (\cite{mr92},
\cite{ds10}).  Because the known number of symbiotic stars (less than 200,
\cite{b00}) is a tiny fraction of the total number expected (two to
three orders of magnitudes larger), we embarked on a systematic search
for symbiotic systems in the Galaxy. The search takes advantage of
IPHAS, the \ha\ survey of the northern Galactic plane (\cite{d05}),
and will be extended to the whole Galactic plane and part of the bulge
with its approaching southern counterpart (VPHAS+).  So far, thirteen
new symbiotic stars have been discovered (Corradi et al. 2008, 2010a,
2010b, \cite{c10c}), which more than doubles the number of systems
previously known in the area covered by IPHAS.  We present here a new
detection, \iphas\ ($l$=90\deg.19 $b$=+3\deg.11), and the study of its
spectroscopic and photometric properties. The source was previously
known to be an \ha-emitting star and was catalogued as HBHA 5202-01 by
\cite{kw99}.
%In the following we present optical spectroscopy and photometric
%observations that show that it is a D-type symbiotic system with a
%carbon star.

\section{Observational data}

\subsection{Spectroscopy}

A 30~min spectrum of \iphas\ was obtained at the 2.5m Isaac Newton
Telescope (INT) on La Palma, Spain, on 1 August 2006 during
recommissioning of the IDS spectrograph.  The R300V grating was used
with a slit width of 1.1~arcsec. This results in a spectral resolution
of 4.8~\AA\ with a dispersion of 1.9~\AA~pix$^{-1}$ over a spectral
range from 4250 to 8300~\AA.  A more recent spectrum was obtained on
19 September 2010 at the 2.1m telescope of San Pedro Martir (SPM),
Mexico, with the Boller \& Chivens spectrograph.  The 400 line
mm$^{-1}$ grating was used combined with the OG550 filter to
remove second order contamination, and with a 2$''$.5 wide slit,
providing a dispersion of 1.9~\AA~pix$^{-1}$, a resolution of 5.5~\AA,
and a spectral coverage from 5850 to 9700~\AA.  Total exposure time
was 60 min. Flux calibration was determined by observing
spectrophotometric standard stars from \cite{o90} immediately after the
target on each night. For the SPM spectrum, the flux calibration is
uncertain above 9200~\AA.  In addition, both spectra suffer from
fringing longward of \ha, which in the SPM spectrum was partially
corrected by lamp flat fields taken during the night.
 
An additional spectrum was obtained on 26 December 2010 at the 4.2m
William Herschel Telescope (WHT) on La Palma. In the red arm of the
ISIS spectrograph, grating R158R was used combined with a 1$"$
wide slit, providing a dispersion of 1.8~\AA~pix$^{-1}$, a resolution
of 6.5~\AA, and a spectral coverage from 5400 to $\sim$10000~\AA. The
exposure time was 20 min. This spectrum was obtained on a cloudy night, 
but under excellent seeing (0$''$.7): for this reason, in the
following it will be only used to investigate the spatial profile and
velocity structure of emission lines.

A high-resolution spectrum around \ha\ was obtained with the 2.1m
telescope at SPM using the Mezcal spectrograph on 21 September 2010.
The 2$''$ wide slit provided a resolution of 0.25~\AA\ with a
dispersion of 0.06~\AA~pix$^{-1}$. Spectral coverage was from 6531 to
6593~\AA, and the exposure time was 90 min.

\subsection{Photometry}

Two IPHAS photometric measurements, both obtained on 10 November
2003, are available. The IPHAS (SDSS) magnitudes of the source at that
time, listed in Table~\ref{T-phot}, indicated the presence of a red
star (\ri=2.0) with a strong \ha\ line in emission (\rha=2.2, cf. the
colour-colour diagram of \cite{c08}). This drew our attention to
the object. Because the global photometric calibration of IPHAS is not
available yet, the error on these magnitudes is conservatively taken
as 0.1 mag.

Seven years later, from November 2010 to January 2011, we obtained new
photometric measurements using the GAPC 0.7m Ritchey-Chr\'etien
telescope at La Polse di Cougnes, Udine, Italy. It is equipped with an
Apogee ALTA U9000 CCD Camera 3056$\times$3056 array, with 12 $\mu$m
pixels $\equiv$ 0.44$^{\prime\prime}$~pix$^{-1}$ and field of view of
22$^\prime$$\times$22$^\prime$.  The instrumental magnitudes, obtained
through R and I Cousins filters, were placed on the photometric system
of the IPHAS survey by determining the transformation colour equations
using fifteen field stars that span a wide range of IPHAS colours that
well encompass those of our programme star.  The new photometric
points are also listed in Table~\ref{T-phot}.  Quoted errors are the
quadratic sum of the Poissonian and colour transformation errors.
%on 9 Nov 2010, we have obtained a new
%photometric measure in the SDSS $i$ band using the Camelot camera at
%the 80cm IAC telescope in Tenerife, Spain, finding a brightening of 
%2.8$\pm$0.1~mag with respect to the 2003 IPHAS measurement. 
%($i$=14.05$\pm$0.05).

\iphas\ was detected by the near-infrared 2MASS survey in 1999.
%2MASS magnitudes are J=13.49, H=10.50, and =8.16~mag. 
We obtained new $J$, $H$ and $K_s$ magnitudes on 24 November 2010
with the 1.5m Carlos Sanchez Telescope (TCS) and the CAIN--III near-IR
camera in Tenerife, Spain. The
photometric zero-points and colour equations were determined by 
exploiting the 2MASS stars that are present in the field.  The
near-IR magnitudes are also listed in Table~\ref{T-phot}.
%Near-IR observations ($JHKs$) have been obtained with the Telescopio
%Carlos Sanchez (TCS), at the Observatorio del Teide (OT) of the
%Instituto de Astrof\'{i}sica de Canarias (IAC), on 24$^{th}$ Nov 2010.
%The telescope is equipped with \textit{CAIN-III}, an infrared camera
%whose detector is composed of a mosaic of 256x256 HgCdTe photovoltaic
%elements.  The pixel's physical size is 40 \mu m; in the so-called
%"wide" configuration, it corresponds to a scale of 1"/pix, namely to a
%field of view of 4.2'x4.2'.  The preliminary data reduction has been
%performed using \textit{CAINDR} (
%http://www.iac.es/telescopes/cain/reduc/caindr.html), an IRAF external
%package specifically developed by the TCS staff for dealing with **
%\textit{CAIN-III}** data. The calibration of photometric zero-points
%and colour equations has then been carried out by exploiting the 2MASS
%stars present in the field.
The source was also detected at longer wavelengths.  It is catalogued 
%The Infrared Astronomical Satellite catalogued it 
as IRAS 20570+5021, and the corresponding flux densities are 
F$_{12}$=6.06, F$_{25}$=3.26, F$_{60}$$<$0.81, and F$_{100}$$<$8.69
Jy. The Midcourse Space Experiment (MSX) provides the following flux
densities: F$_{8.28}$=5.43$\pm$0.22~Jy, F$_{12.13}$=4.88$\pm$0.25~Jy,
F$_{14.65}$=3.75$\pm$0.23~Jy, and F$_{21.34}$=2.45$\pm$0.17~Jy.  It is
also reported in the AKARI/IRC All-Sky Survey Point Source Catalogue 
with the following flux densities: F$_9$=4.78$\pm$0.26~Jy and
F$_{18}$=3.05$\pm$0.19~Jy.

\begin{table}
%\centering
  \caption{Optical and near--IR photometric data.}
\begin{tabular}{ccccrl}
\hline\hline\\[-8pt]
Date       & HJD$-$2450000& Source       & Band      &  Mag   & Error  \\
\hline\\[-5pt]
2003-11-10 & 2954.39  & IPHAS$^\star$ &H$\alpha$& 16.63  & 0.10$\ddag$\\[3pt]
2003-11-10 & 2954.39 & IPHAS$^\star$ & $r'$    & 18.87  & 0.10$\ddag$\\
2010-11-23 & 5524.38 & GAPC         & $r'$    & 16.76  & 0.02      \\
2010-11-24 & 5525.34 & GAPC         & $r'$    & 16.83  & 0.02      \\
2010-11-29 & 5530.41 & GAPC         & $r'$    & 16.79  & 0.02      \\
2010-12-04 & 5535.40 & GAPC         & $r'$    & 16.80  & 0.02      \\
2010-12-09 & 5540.37 & GAPC         & $r'$    & 16.84  & 0.02      \\
2010-12-11 & 5542.39 & GAPC         & $r'$    & 16.77  & 0.01      \\
2010-12-29 & 5560.33 & GAPC         & $r'$    & 16.83  & 0.01      \\
2011-01-13 & 5575.26 & GAPC         & $r'$    & 17.00  & 0.02      \\
2011-01-17 & 5579.26 & GAPC         & $r'$    & 17.07  & 0.01      \\
2011-01-22 & 5584.26 & GAPC         & $r'$    & 17.02  & 0.01      \\[3pt]
2003-11-10 & 2954.39 & IPHAS$^\star$ & $i'$    & 16.88 & 0.10$\ddag$\\
2010-11-23 & 5524.38 & GAPC         & $i'$    & 14.11  & 0.01     \\
2010-11-24 & 5525.34 & GAPC         & $i'$    & 14.12  & 0.01     \\
2010-11-29 & 5530.41 & GAPC         & $i'$    & 14.13  & 0.01     \\
2010-12-04 & 5535.40 & GAPC         & $i'$    & 14.15  & 0.01     \\
2010-12-09 & 5540.37 & GAPC         & $i'$    & 14.12  & 0.01     \\
2010-12-11 & 5542.39 & GAPC         & $i'$    & 14.10  & 0.01      \\   
2010-12-29 & 5560.33 & GAPC         & $i'$    & 14.15  & 0.01      \\
2011-01-13 & 5575.26 & GAPC         & $i'$    & 14.26  & 0.01      \\
2011-01-17 & 5579.26 & GAPC         & $i'$    & 14.30  & 0.01      \\
2011-01-22 & 5584.26 & GAPC         & $i'$    & 14.30  & 0.01      \\[3pt]
1999-06-21 & 1350.90 & 2MASS$\dag$  & $J$     & 13.49  &  0.02    \\
2010-11-24 & 5525.36 & TCS          & $J$     &  9.74  &  0.18    \\[3pt]
1999-06-21 & 1350.90 & 2MASS$\dag$  & $H$     & 10.50  &  0.02    \\
2010-11-24 & 5525.36 & TCS          & $H$     &  7.64  &  0.10    \\[3pt]
1999-06-21 & 1350.90 & 2MASS$\dag$  & $K_s$   &  8.16  &  0.01    \\
2010-11-24 & 5525.36 & TCS          & $K_s$   &  6.08  &  0.15    \\
\hline\\ 
\end{tabular}
\newline $^\star$ All optical data 
are tied to this photometric system.\\
$\dag$ All near-IR data are tied to this photometric system.\\
$\ddag$ Error includes uncertainty of IPHAS global calibration (see text).
\label{T-phot}
\end{table}

\begin{figure*}[!ht]
\centering
\includegraphics[width=18.0cm]{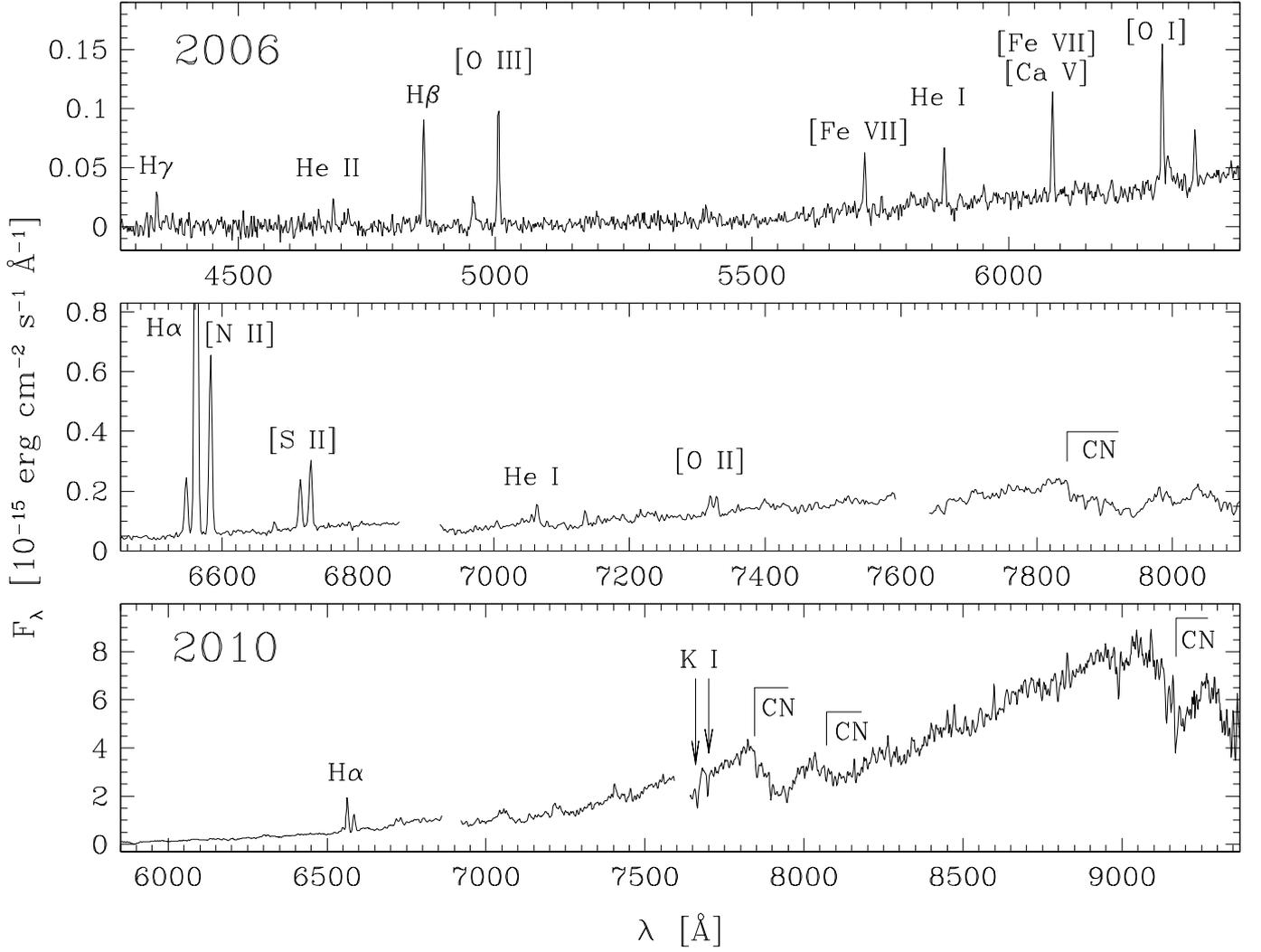}
\caption{2006 (upper and middle rows) and 2010 (bottom panel)
  spectra of \iphas. Only the regions with reliable flux
  calibration are plotted.  Note the different x- and y-scales in the
  various panels. The regions around 6870~\AA\ and 7600~\AA\ containing
  the O$_2$ atmospheric absorption bands are not plotted.}
\label{F-spectra}
\end{figure*}

\section{Analysis}

\subsection{Emission-line spectrum}

The 2006 and 2010 low-resolution spectra of the source are presented in
Figure~\ref{F-spectra}.  In 2006, the spectrum showed narrow emission
lines from low (e.g. \oi, \feii, \sii) to high (\oiii, \heii, \cav,
\fevii) ionization species. They are listed in
Table~\ref{T-emlines2006}: observed line fluxes F$_{obs}$ are given
relative to F$_{H\beta}$=100, and the estimated \hb\ flux is
4.9$\times$$10^{-16}$~erg~cm$^{-2}$~s$^{-1}$. The error on the
absolute flux of \hb\ is 20\%, caused by a lack of precise absolute
flux calibration for these data. The errors in the quoted fluxes
(relative to \hb) for the other lines are $\sim$5\%\ for lines
stronger than 0.5\,F$_{H\beta}$, and are larger for fainter lines.
The emission-line spectrum is for instance similar to that of the
recently discovered D-type symbiotic star \object{K~3-22}
(\cite{c10c}), but with stronger low-excitation forbidden lines
such as \nii, \sii, or \oi.

For any reddening value and standard interstellar extinction laws, the
Balmer \ha, \hb, and \hg\ lines do not show the theoretical ratios
expected for low-density nebulae ($N_e$$\le$$10^7$~cm$^{-3}$,
\cite{b71}, \cite{km78}).  For higher densities, the intrinsic line
ratios become a complex function of $N_e$, $T_e$, and optical depth
(\cite{d80}).  
A high \ha/\hb\ ratio as observed in \iphas \ (=27) can be produced
for large optical depths in the Balmer lines
($\tau_{H\alpha}$$\ge$100) and/or relatively high densities
($10^6$$\le$$N_e$$\le$$10^{10}$~cm$^{-3}$). At higher densities the
ratio decreases again (\cite{d80}). The \hg/\hb\ ratio has a complex
behaviour, but remains at a level closer to the low-density case for
($N_e$$<$$10^{8}$~cm$^{-3}$) unless $\tau_{H\alpha}$ is very high.
Without reliable density, temperature, and optical depth
indicators for \iphas, the estimate of accurate intrinsic ratios for
the Balmer lines, and hence the calculation of the reddening, is
precluded.  Even so, if we take $N_e$$<$$10^{10}$~cm$^{-3}$ for the
region where the bulk of the \hi\ is emitted, the observed
\hg/\hb\ ratio (=0.3) provides an upper limit of 4~mag to the reddening
in the V band, using the \cite{f04} extinction law.  We take this
figure as the upper limit to the interstellar reddening of \iphas.
%This value is consistent with the asymptotic value of the interstellar
%reddening in the direction toward \iphas\ according to the IPHAS
%extinction maps (Sale et al, in prepration), and the ones by Drimmel
%This is also supported by the lack of detection of a strong
%\nii 5755 line in our spectrum.

\begin{table}
%\centering
  \caption{Emission line fluxes observed in 2006, relative to F$_{H\beta}$=100.}
\begin{tabular}{llrllr}
\hline\hline
Ident.       & $\lambda [\AA]$ & F$_{obs}$\phantom{pip} & \phantom{pip}
Ident.       & $\lambda [\AA]$ & F$_{obs}$ \\
\hline\\[-4pt]
\hi\         & 4340.5 &   30 \phantom{pip} & \phantom{pip}\hi\         & 6562.8 & 2685 \\ 
\heii\       & 4685.7 &   21 \phantom{pip} & \phantom{pip}\nii         & 6583.4 &  682 \\ 
\hi\         & 4861.4 &  100 \phantom{pip} & \phantom{pip}\hei\        & 6678.2 &   30 \\ 
\oiii\       & 4958.9 &   35 \phantom{pip} & \phantom{pip}\sii\        & 6716.4 &  184 \\ 
\oiii\       & 5006.8 &  108 \phantom{pip} & \phantom{pip}\sii\        & 6730.8 &  253 \\ 
\heii\       & 5411.5 &   10 \phantom{pip} & \phantom{pip}\arv\        & 7005.4 &   20 \\ 
\fevii\      & 5720.7 &   49 \phantom{pip} & \phantom{pip}?            & 7056   &   27 \\ 
\hei\        & 5875.6 &   48 \phantom{pip} & \phantom{pip}\hei\        & 7065   &   67 \\ 
\cav$^\star$  & 6086.4 &   93 \phantom{pip} & \phantom{pip}\ariii\      & 7135.8 &   54 \\ 
\oi\         & 6300.3 &  126 \phantom{pip} & \phantom{pip}\feii\       & 7155.1 &   21 \\ 
\siii\       & 6312.1 &   49 \phantom{pip} & \phantom{pip}\oii\        & 7319   &   98 \\ 
\oi          & 6363.8 &   40 \phantom{pip} & \phantom{pip}\oii\        & 7330   &   67 \\ 
\nii         & 6548.1 &  231 \phantom{pip} & \phantom{pip}             &        &      \\
\hline\\[-5pt]                             
\end{tabular}
\newline$^\star$with contribution of \fevii\ 6086.3 at 10\%\ level.\\
%$:$ uncertain flux because of blending.
\label{T-emlines2006}
\end{table}

\begin{figure}
\centering
\includegraphics[width=9.0cm]{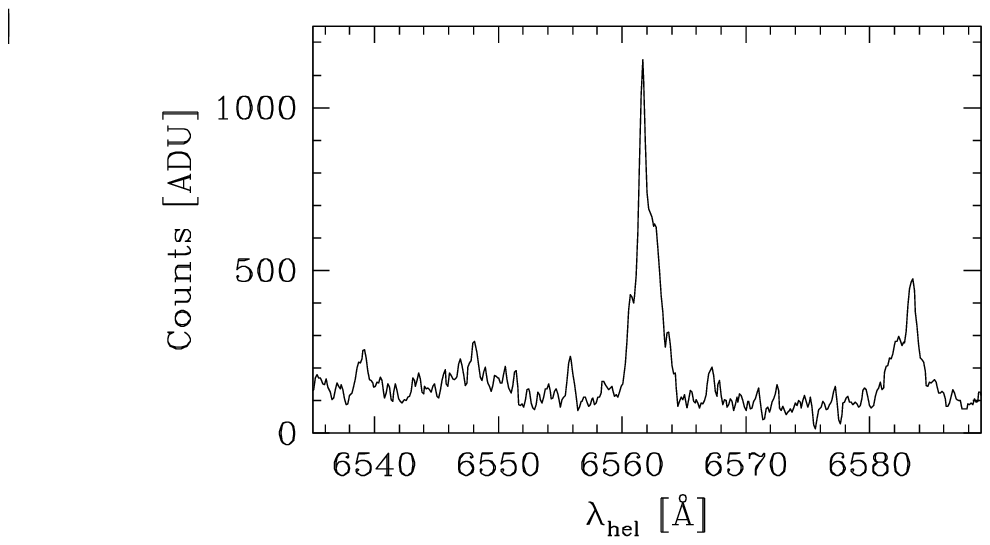}
\caption{High-resolution spectrum around \ha\ of November 2010.}
\label{F-haprof}
\end{figure}

Figure~\ref{F-haprof} shows the profiles of the lines around \ha.  The
signal-to-noise ratio is limited, but clearly both the \ha\ and
\nii6583 line profiles are complex, with multiple components that make
the determination of the systemic velocity $v_{sys}$ of
\iphas\ uncertain. A single-Gauss fit to \ha\ and the mean value of
the two-Gaussian fit of the \nii6583 line would indicate a value
$v_{sys}=-27\pm10$~\kms, corrected to the local standard of rest, but
this number should be taken with caution.  Should the system
participate in the general circular rotation around the Galactic
centre, this systemic velocity would imply a distance of
$3.0\pm0.8$~kpc (\cite{bb93}).
%$4.5^{+2.5}_{-3.0}$~kpc, the relatively large error being mainly due
%to the velocity dispersion ellipsoids of stars in the Galaxy
%(\cite{nordstrom}).
The \ha\ profile has a full-width-at-zero intensity of 180~\kms. Its
articulated shape, the absence of extended wings, and the presence of
a relatively strong \nii\ emission, is visible in several D-type
symbiotic stars such as R~Aqr or He~2-104. This profile corresponds to
type D-1 in the classification scheme of \cite{vw93}.

\subsection{Resolving the nebular component}

\begin{figure}[!h]
\flushright
\includegraphics[width=7.75cm]{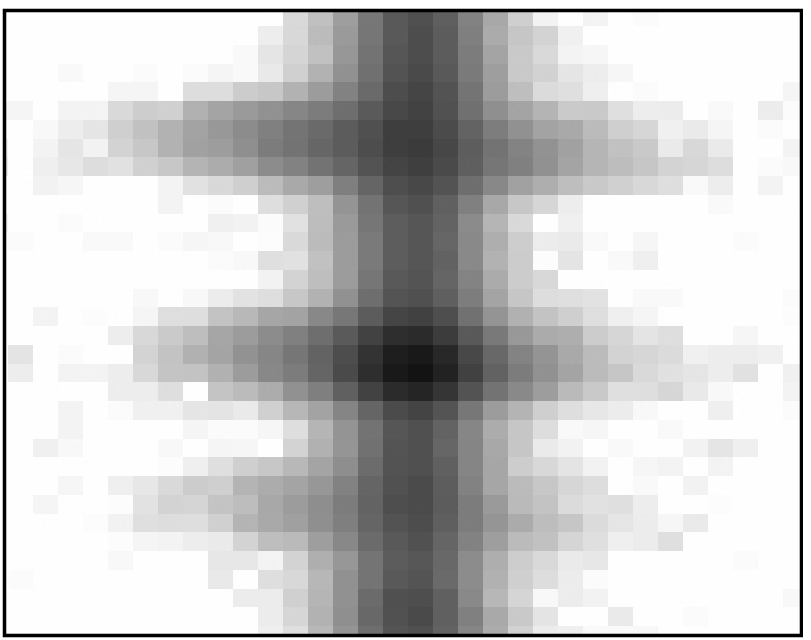}
\includegraphics[width=9.0cm]{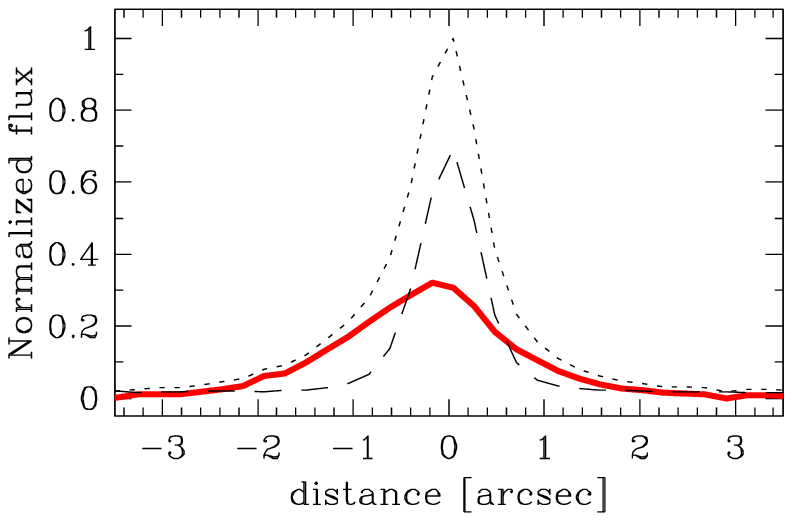}
\caption{{\bf Top}: zoom of the long-slit WHT spectrum around
  \ha\ (middle spectral line) and the \nii\ doublet. The wavelength
  increases from bottom to top. The slit was positioned at P.A.=97\deg,
  and East is towards the left. The size of the box in the spatial
  direction is as in the bottom panel.  {\bf Bottom}: spatial profile
  of the \nii6583 line.  The observed profile is indicated by the
  dotted line, the adjacent stellar continuum by the dashed line, and
  the continuum-subtracted \nii\ profile by the thick (red) solid
  line.}
\label{F-profnii}
\end{figure}

Most of the forbidden lines observed in the spectrum of \iphas\ have
critical densities for collisional quenching of the order of
$N_e$$\sim$$10^{6}$~cm$^{-3}$. They are likely to form in regions more
extended and less dense than the nebular/stellar core where most of
the \hi\ is emitted (see e.g. \cite{cs03}). Indeed, the excellent
seeing of the 2010 WHT spectrum allows us to resolve the inner nebula.
In this spectrum, obtained with the slit orientated at a position angle
97\deg, the full-width-at-half maximum (FWHM) of the source along the
spatial direction is 0$''$.7 at wavelengths dominated by the stellar
continuum (line-free).  The spatial profile of \ha\ and more clearly
of the nearby \nii\ and \sii\ doublets is instead broader with
non-Gaussian extended wings, indicating that the gas emitting region
is partly resolved (Figure~\ref{F-profnii}, top).  This is best seen
in the \nii6583 line, because of the relatively weaker contribution
from the unresolved core.  The bottom panel of Figure~\ref{F-profnii}
shows the observed spatial profile of \nii6583 (dotted line), the
profile of the continuum at adjacent wavelengths (dashed line), and
the continuum-subtracted \nii\ profile (thick red solid line). The
latter is fairly well fitted by a Gaussian of 1$''$.9 FWHM, or 1$''$.8
after seeing deconvolution. This figure would underestimate the true
nebular diameter by a factor from 1.2 to 1.6 for standard morphologies
(thin or thick shells, disc, see \cite{vh00}). Taking an intermediate
correction factor, the resulting size of 2$''$.5 corresponds to a
linear diameter $\ge$0.05~pc for a distance larger than 4~kpc (see
Sect.~3.4). Such a compact nebula might be similar to those around the
symbiotic Miras \object{Hen~2-147} (\cite{mp93}, \cite{s07}) or
\object{RX~Pup} (\cite{cs03}).

The seeing in the original IPHAS images (1$''$.2) and the width of the
\ha\ filter used (95~\AA), which increases the relative contamination
from the point-like core, prevented a clear detection of the extended
nebula in those data. A careful analysis of the point-spread function
in the images shows that the source is indeed marginally resolved,
with weak evidence for elongation roughly along the east-west
direction.

In the WHT spectrum, a slight tilt of the emission line is also
visible. The brighter eastern side of \ha, \nii\ and \sii\ is
redshifted with respect to the fainter western side by
1.0$\pm$0.3~\AA, or 45$\pm$15~\kms.  This corresponds, at least
partially, to the velocity components in the spatially-unresolved \nii\ profile
of the high-resolution spectrum of Figure~\ref{F-haprof}.  The
observed velocity structure indicates that the nebula is not
spherical: an axisymmetrical or ring-like morphology as commonly
observed in symbiotic stars (\cite{c03}) is suggested.  Imaging at
sub-arcsec resolution through a narrow \nii\ filter is needed to gain
more information about the extension and morphology of the
circumstellar nebula of \iphas.

\subsection{The stellar continuum}

In the 2006 spectrum, a red continuum is observed. It shows the
absorption bands that are typical of a carbon-rich star, in particular
the CN band starting at 7845~\AA.  In 2010, the stellar continuum has
significantly brightened, by roughly the same amount as determined
comparing the 2003 and 2010 photometry.
%indicating that brightening mainly occurred after 2006.  
%The faintest emission lines are now vanished in the strong
%continuum in the red.  
Within the errors, the \ha\ line flux has not varied from 2006 to 2010,
but the flux of the \nii\ and \sii\ doublets is about twice as large.
%No information is available about possible variations of bluer lines.

The redder setup of the 2010 spectrum includes spectral features
sensitive to the stellar temperature in carbon stars. The strong
\ki\ absorption lines at 7665 and 7699, together with the absence of
the \caii\ infrared triplet, and in particular of the \caii~8662 line,
puts the object at the cool end of the carbon star sequence
(\cite{ri71}), with a temperature $T_{eff}$$\sim$2000~K
(\cite{be01}). This corresponds to spectroscopic class C9
(\cite{ri71}), or photometric class CV7 (\cite{k99}).  A large 
percentage of sources with these characteristics are variables at the
tip of the asymptotic giant branch (AGB): i.e. Miras or semi-regular
variables (\cite{k99}).
%\footnote{According to Whitelock et al. (1997), the distinction
%between Miras and SRs in carbon stars is not very clear. We will
%therefore use only the term ``carbon Mira'' in the following.}.

\subsection{Photometry}

The photometric data available before the present season were
obtained in 1999 in the near-IR ($J$, $H$, and $K_s$ bands, 2MASS) and
in 2003 in the red ($r'$ and $i'$, IPHAS). A comparison with the
present measurements shows that the brightening of the source has been
significant, from 2.1~mag ($K_s$ and $r'$) up to 3.8~mag in
$J$\footnote{This is likely the band in which the cool giant continuum
  is less affected by the combined contribution of the hot companion,
  ionized nebula, and circumstellar dust emission.}
(cf. Table~\ref{T-phot}). 
While the \ri\ colour has become redder, in the near-IR \iphas\ is now
substantially bluer than eleven years ago.  As shown in the previous
sections, the main cause of the variation is the increase in
brightness of the carbon star. In the optical, the contribution of the
ionized nebula is much larger in the $r'$ band than in $i'$: assuming
that the cool star has brightened but the nebula
did not vary much would explain the redder \ri\ colour observed in
2010.  Sticking to the near-IR data where the nebular contribution is
negligible, and correcting for the upper limit of the interstellar
extinction estimated above, we obtain that in 2010 \jho$>$1.8 and
\hko$>$1.3.  

These colours are typical of carbon Miras (see Figure~1 in
\cite{w00}).  These stars, and in particular those with
moderate-to-thick dust shells, display large-amplitude, long-term
photometric variations in addition to the conspicuous light modulation
caused by pulsations (\cite{w97}).  The long-term variations are often
associated with periodic or erratic dust formation/obscuration episodes
(\cite{w00}), but orbital effects in a binary system have also been
invoked, e.g., for V~Hya (\cite{knapp99}). The
combination of pulsations and long-term variability observed in other
carbon Miras is indeed consistent with the large photometric changes
of \iphas\ in the last decade.  Its even more extreme colours in 1999, 
\jho$>$2.7 and \hko$>$2.1, are characteristic of carbon Miras
with thick dust shells like \object{IRC+10216}, \object{V688~Mon}, or
\object{IZ~Peg}.  Such a strong near-IR excess is ascribed to the
combination of extinction and thermal emission from dust grains in
circumstellar shells of a substantial optical depth (\cite{k99}).
%At least to a first approximation, the redder the near-IR colours, the
%thicker the circumstellar dust shell (\cite{w97}).
The [12]-[25] IRAS colour of the source is also consistent with a
classification as a carbon Mira (\cite{w97}).  

\begin{figure}[!ht]
\centering
\includegraphics[width=9.0cm]{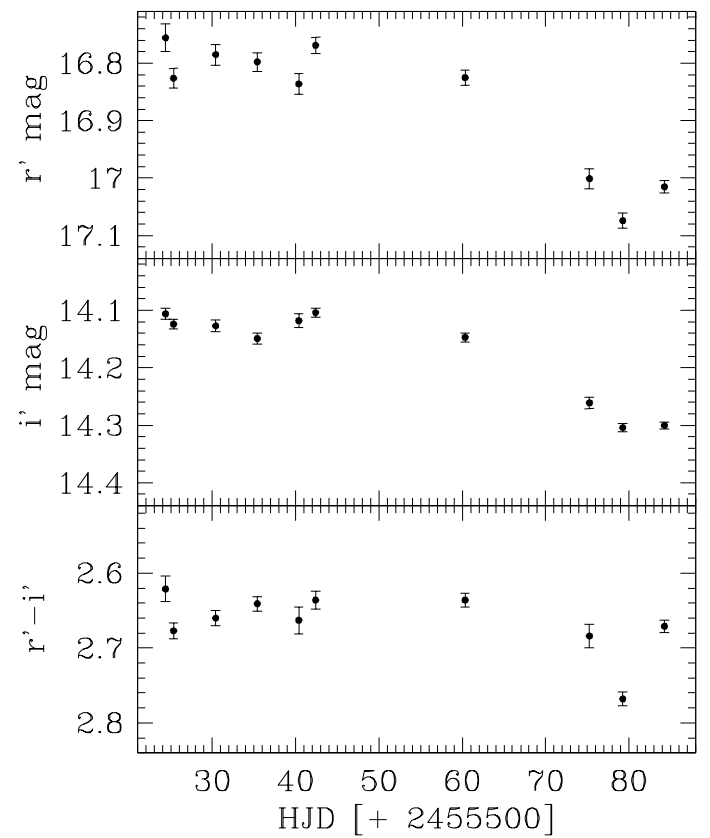}
\caption{Photometric behaviour of \iphas\ from November 2010 to January 2011.}
\label{F-plotgapc}
\end{figure}

Indeed, a slow variability was detected over a period of 60 days
between November 2010 and January 2011 (Table~\ref{T-phot} and
Figure~\ref{F-plotgapc}). The observed luminosity decrease might be
caused by the Mira pulsations, but it is premature to draw conclusions
based on this two-month monitoring. Combining these data with the 2003
IPHAS measurements and with the available near-IR photometry should be
done with caution, given the above-mentioned additional variability
related to dust formation events, which are common in these stars and
are indicated in \iphas\ by the variations of its near-IR magnitude
and colour ($\Delta $J=3.8~mag, $\Delta$(\jk)=1.6~mag), which are
larger than expected from Mira pulsation only (see Tables 2 and 3 of
\cite{w06}).

With the hypothesis that the cool giant is indeed a Mira, some rough
limits to the distance of \iphas\ can be derived.  We assume that the
shape of the light curve in Figure~\ref{F-plotgapc} indicates that the
star was near maximum at the end of 2010 ($K_{max}=6.1$~mag). In
addition, the data in Table 3 of \cite{w06} show a correlation between
the pulsational period and amplitude of the light curve in the
$K$-band for carbon Miras.  This relation provides a way to estimate
the mean magnitude of the carbon star of \iphas\ throughout a
pulsational cycle. For a period $P$=300~days (unlike given the slow
variability in Figure~\ref{F-plotgapc}), a mean magnitude
$K_{mean}=6.3$~mag is derived. For $P$=800~days, $K_{mean}=7.1$~mag.
A reddening $A_K$$\sim$0.8~mag of the carbon Mira at the end of 2010
was estimated by adopting from Bergeat, Knapik \& Rutily (2001) the
intrinsic mean \jk\ colour of carbon stars with $T_{eff}$$\sim$2000~K
(see Sect. 3.3). The application of the period-luminosity relationship
of \cite{w08} results in a distance of 4~kpc if $P$=300~days, and
11~kpc if $P$=800~days.

\section{Conclusions}

Our observations indicate that \iphas\ is a new symbiotic star of the
D-type with a carbon-rich cool giant, likely a carbon Mira.  Only two
other Galactic symbiotic stars are known to have a carbon Mira.  They
are \object{SS~38} and \object{AS~210} (\cite{g09}). Both show an
emission-line spectrum (including the 6825~\AA\ Raman
scattered feature) stronger than in \iphas, and have warmer cool giants
(\object{SS~38} being the most similar).  The other candidates in the
catalogue of Belczy\'nski et al. (2000), \object{StH$\alpha$-55}
(\cite{mu08}) and \object{UV~Aur} (e.g. \cite{h09}), were instead
shown not to be symbiotic stars. New data are needed to confirm the
carbon Mira V335~Vul as a symbiotic star.

Summarising, \iphas\ seems to belong to a rare group of stars in the
Galaxy: carbon Miras in interacting binary systems. Its 2000~K giant
with a thick circumstellar dust shell locates it at the tip of the
AGB, about to loose its entire envelope. The object and its resolved
ionized nebula can provide valuable information about the effects of a
close companion on aspects like the pulsational properties of carbon
Miras, the cause of long-term periodic photometric variations (how
many of them are related to binarity?), the mass loss geometry and its
timescales (highly relevant to the formation of asymmetric outflows in
evolved stars and related nebulae). This is particularly important
given the limited information available on carbon Miras in general
(cf. \cite{k99}).  To reach these goals, multi-wavelength photometric
monitoring should continue to confirm the presence of a Mira variable
in \iphas, determine its pulsational period (and hence the distance of
the system using the period-luminosity relation), separate pulsations
from long-term variability caused by dust obscuration events, and
compare the main properties of the system with other binary and single
carbon Miras.

Extended nebulae are common in D-type symbiotic Miras (\cite{c03}),
but they were not previously detected around carbon-rich symbiotic
stars.  These nebulae are thought to be mainly composed of red giant
wind that is not accreted by the white dwarf companion. A study of the
properties and chemical composition of the nebula of \iphas\ would be
of great value to test its origin, and perhaps reveal chemical changes
in the carbon star in the last $\sim$$10^4$~years.  Finally, the
parameters of the binary system (distance, stellar masses and
luminosities, orbital period) should also be determined by a variety
of methods. With a lot of patience, though, given the long orbital,
pulsational, and dust-evolution timescales of these systems.

%\begin{figure}[!ht]
%\centering
%%\includegraphics[width=8.0cm]{fig_pap3.eps}
%%%BoundingBox: 78 521 300 693
%\caption{Interstellar extinction vs. distance for the sightline toward
%  \iphas. The broad (red) line indicates the range of possible distance
%  and extinction values determined from modelling the 2006 spectrum: its
%  left end corresponds to an M4.0~III type and E(B-V)=1.69, and its right
%  end to M6.0~III and E(B-V)=1.55.}
%\label{F-extdist}
%\end{figure}

\begin{acknowledgements}
RLMC and AM acknowledge funding from the Spanish AYA2007-66804 grant.
RA acknowledges a grant from the FONDECYT Project N. 3100029.
This paper makes use of data obtained as part of the INT Photometric
\ha\ Survey of the Northern Galactic Plane (IPHAS) carried out at the
Isaac Newton Telescope (INT). All IPHAS data are processed by the
Cambridge Astronomical Survey Unit, at the Institute of Astronomy in
Cambridge.
%This publication also makes use of data products from the Two Micron All
%Sky Survey, which is a joint project of the University of
%Massachusetts and the Infrared Processing and Analysis
%Center/California Institute of Technology, funded by the National
%Aeronautics and Space Administration and the National Science
%Foundation.
\end{acknowledgements}

\end{document}